\documentclass[12pt]{iopart}

\usepackage[pdftex]{graphicx}
\usepackage{dcolumn}
\usepackage{bm}

\usepackage[T1]{fontenc}
\usepackage[latin1]{inputenc}
\usepackage[english]{babel}
\usepackage{amssymb}
\usepackage{amstext}
\usepackage{relsize}
\usepackage{subfigure}
\usepackage{color}
\usepackage{ulem}

\begin{document}

\title[Density dependent tunneling in the extended Bose-Hubbard model]{Density dependent tunneling in the extended Bose-Hubbard model}

\author{Micha\l{} Maik$^{1,2}$, Philipp Hauke$^{2,3}$, Omjyoti Dutta$^{1,2}$, Maciej Lewenstein$^{2,4}$ and  Jakub Zakrzewski$^{1,5}$ }

\address{$^1$ Instytut Fizyki imienia Mariana Smoluchowskiego,
Uniwersytet Jagiello\'nski, ulica Reymonta 4, PL-30-059 Krak\'ow, Poland}

\address{$^2$ ICFO -- Institut de Ci\`encies Fot\`oniques,
Mediterranean Technology Park, E-08860 Castelldefels (Barcelona), Spain}

\address{$^3$ Institute for Quantum Optics and Quantum Information of the Austrian Academy of Sciences, A-6020 Innsbruck, Austria}

\address{$^4$ ICREA -- Instituci\'{o} Catalana de Recerca i Estudis Avan\c{c}ats, E-08010 Barcelona, Spain}

\address{$^5$ Mark Kac Complex Systems Research Center,
Uniwersytet Jagiello\'nski, Krak\'ow, Poland}

\ead{michal.maik@uj.edu.pl}

\date{\today}

\begin{abstract}

Recently, it has become apparent that, when the interactions between polar molecules in optical lattices becomes strong, the conventional description using the extended Hubbard model has to be modified by additional terms, in particular a density-dependent tunneling term. 
We investigate here the influence of this term on the ground-state phase diagrams of the two dimensional extended Bose--Hubbard model. Using Quantum Monte Carlo simulations, we investigate the changes of the superfluid, supersolid, and phase-separated parameter regions in the phase diagram of the system. By studying the interplay of the density-dependent hopping with the usual on-site interaction $U$ and nearest-neighbor repulsion $V$, we show that the ground-state phase diagrams differ significantly from the ones that are expected from the standard extended Bose--Hubbard model.  However we find no indication of pair-superfluid behaviour in this two dimensional square lattice study in contrast to the one-dimensional case.

\end{abstract}


%
\maketitle


\section{Introduction}

In the last decade, the physics of ultra-cold atoms in optical-lattice potentials has undergone extensive developments due to the extreme controllability and versality of the realizable many-body systems (for recent reviews see \cite{Lewenstein07,Bloch08}). 
The tight-binding description predicted in 1998 \cite{Jaksch98}, termed Bose--Hubbard model (BHM) for bosonic atoms with contact $s$-wave interactions, was soon after verified via the  experimental observation of the superfluid (SF) -- Mott insulator (MI) 
transition \cite{Greiner02}. 
For particles interacting via a long-range (e.g., dipole-dipole) potential, the original model has to be modified, typically including a density--density interaction between different sites. The simplest approximation, taking into account only the interaction between nearest neighbours, is termed the extended Bose-Hubbard 
model (EBHM). 
As compared to the BHM, the extended model allows for the existence of novel quantum phases such as checkerboard solids, supersolid phases \cite{Wessel05b, Heidarian05, Melko05, Sengupta05, Batr, Cap}, exotic Haldane insulators \cite{Dalla}, and more. 

Recently, however, it has been realized that even in the simpler case of contact $s$-wave interactions, in certain parameter regimes, carefully performed tight-binding approximations lead to an additional correlated tunneling term in the resulting microscopic description. This term, known in the case of fermions as bond-charge contribution \cite{hirsch89}, is even more important for bosons \cite{Dutta2011, Luhmann2012,Lacki2013}. 
It is found that such tunneling terms along with the effect of higher bands can 
provide an explanation \cite{Luhmann2012,Mering2011} of the unexpected shift in the MI--SF transition point for Bose--Fermi \cite{Ospelkaus2006, G\"unter06} and Bose-Bose mixtures \cite{Catani08} as well as shifts in absorption spectra for bosons in optical lattices \cite{Lacki2013}.

One may expect that similar bond-charge (or density-dependent tunneling) effects may play a similarly important role in the presence of dipolar interactions. 
This assumption has been verified by some of us \cite{Sowinski} in a recent study, where it has been shown that the additional terms in the Hamiltonian may destroy some insulating phases and can create novel pair-superfluid states. 
That study \cite{Sowinski} has been restricted to a one-dimensional (1D) model due to the numerical methods used. 
Here, we use Quantum Monte Carlo (QMC) methods to study soft-core dipolar gases trapped in two-dimensional square optical lattices, where we assume a tight confinement in the remaining $z$ direction (which is also the  polarization direction of the dipoles). 
A similar two-dimensional model without density-dependent tunneling terms was analyzed before \cite{Sengupta05}, providing  us with a benchmark against which we may test the importance of density-dependent tunneling.
In Ref.~\cite{Sengupta05}, a supersolid phase was observed in the EBHM at half filling.  
Such a supersolid is characterized by the coexistence of superfluid and crystal-like density--density diagonal long-range order \cite{Wessel05b, Heidarian05, Melko05, Sengupta05, Batr, Cap}. 
Experimental evidence of this counter-intuitive quantum phase is still missing, since the claim of an experimental realization of supersolidity in$\phantom{a}^4$He \cite{Kim2004,Kim2004a} could not be reproduced in later experiments \cite{Kim2012, Kuklov2011}.
As we shall see, in the present model, the sign of the additional tunneling (or, more precisely, the relative sign between the standard tunneling and the density-dependent one) can stabilize or destabilize the supersolid phases. 

\section{The model}

The appropriate tight-binding model to study interacting dipolar bosons occupying the lowest band in a lattice reads
 \cite{Sowinski}
\begin{eqnarray}
\label{eq:EBHM}
H &= &-t\sum\limits_{\langle i,j \rangle}(a^{\dag}_{i}a_{j}+\mathrm{h.c.})+\frac{U}{2}\sum\limits_{i}n_{i}(n_{i}-1)
         +V\sum\limits_{\langle i,j\rangle}n_{i}n_{j} \nonumber \\
    &-&  T\sum\limits_{\langle i,j \rangle}(a^{\dag}_{i}(n_{i}+n_{j})a_{j}+\mathrm{h.c.})
        +P\sum\limits_{\langle i,j \rangle}(a^{\dag}_{i}a^{\dag}_{i}a_ja_{j}+\mathrm{h.c.})
      -\mu\sum\limits_{i}n_{i},
  \label{Ham}
\end{eqnarray}

\noindent where $a^{\dag}_{i}$ $(a_{i})$ is the creation (annihilation) operator of a boson at site $i$ and $n_{i}$ is the number
operator; $t$ is the regular hopping term, $U$ the onsite repulsion, and $\mu$ the chemical potential. We assume a system of dipolar bosons in a 2D square lattice with dipolar moments polarized perpendicularly to the lattice, thus leading to dipole-dipole repulsion.  Then, the present model contains three terms that come from the dipolar interactions, the nearest-neighbor repulsion $V$, the density-dependent hopping $T$, and the  correlated pair tunneling $P$. We restrict here the range of $V$ to the nearest neighbors to allow for a direct comparison with the results of Ref. \cite{Sengupta05} and \cite{Sowinski} Within the standard EBHM, both the $T$ and $P$ terms are neglected. However, the analysis presented in Ref.~\cite{Sowinski} has shown that, although $V$ is typically an order of magnitude larger than both $T$ and $P$, the latter terms cannot be neglected in the presence of strong dipolar interactions.    

The four parameters $U$, $V$, $T$, and $P$ have the same physical origin, namely interactions, and are therefore correlated. However, in the two-dimensional model, changing the trapping frequency in the direction perpendicular to the plane affects quite strongly only the on-site $U$ term (for dipolar as well as for the contact part of the interactions), while leaving the other three parameters practically unaffected \cite{Sowinski}.
Thus, we shall consider $U$ as an independent parameter. 
To facilitate a comparison with earlier works  (e.g., \cite{Sengupta05}) that did not take $T$ tunneling into account we span a similar parameter range for $U$, $V$, and filling fractions. 
The values of $T$, $V$, and $P$  are strongly correlated  as they originate from nearest-neighbour scattering due to long-range interactions.
For a broad range of optical-lattice depths, the parameters $T$ and $V$ are typically related as $V\approx|10T|$.  The absolute value of $P$ is almost another magnitude smaller than $T$ (compare Fig.~1 of \cite{Sowinski}).  Thus, for simplicity, we will set 
$V=|10T|$ in the following and neglect the $P$ term altogether. 
This will allow us to study in depth the effects due to density-dependent tunnelings.  
The previous study \cite{Sowinski} has shown that there is a broad tunability regarding the relationship of the two tunneling parameters $T$ and $t$, allowing a regime where the two hopping terms have opposite signs, and even the exotic situation that $T$ dominates over $t$. 

Whether both hopping terms are of the same or of opposite sign has a major influence on the phases that will appear in the system.  
Generally speaking, when both hopping terms have the same sign, one can expect an increase of the influence of the overall hopping. 
Otherwise, if the signs are opposite, there will be a competition between the two terms. 
Therefore, the influence of the additional density-dependent hopping can be expected to strongly affect  the phase diagram.

To form an intuition about our system, let us give a brief summary of the results from the previous study \cite{Sowinski} of a similar 1D system with both a density-dependent and a pair-hopping term. 
In that study, exact diagonalization (system sizes between $L = 8$ and $L = 16$) and the Multiscale Entanglement Renormalization Ansatz (MERA) (system sizes up to $L=128$) were used to study the phase diagram at zero temperature.  
The results, when both $T$ and $P$ are set to zero, show the existence of three phases. 
At weak interaction, there is a superfluid phase (SF), while at stronger dipolar strength, two 
charge density wave (CDW) phases appear. The CDW phases are characterized by a periodic, crystal-like structure where occupied and empty sites alternate in a checkerboard pattern. 
In the following, we denote cases where the populated sites are occupied by a single atom (two atoms) as CDW I (II).  
In the one-dimensional case of Ref.~\cite{Sowinski}, the two observed CDW phases are a CDW I phase at half filling with a modulation of $|...010101...\rangle$ and a CDW II phase at unit filling with a modulation of $|...020202...\rangle$. 
Now, when the extra terms $T$ and $P$ are incorporated into the Hamiltonian, besides an overall deformation of the phase diagram, there appears also a novel pair-superfluid phase (PSF).  
This more exotic phase is characterized by a finite two-particle NN correlation function $\Phi_{i} =\sum_{\{j\}} \langle a^{\dag}_{j} a^{\dag}_{j} a_{i} a_{i} \rangle$ and a smaller non-vanishing one-particle correlation function $\phi_{i}= \sum_{\{j\}} \langle a^{\dag}_{j} a_{i} \rangle$. 
On the other hand, no supersolid phase has been observed in \cite{Sowinski}. 
In the present study of a 2D lattice, on the contrary, we do observe a supersolid behavior, but we do not find any indications for the existence of a PSF phase. 

\subsection{Considered observables\label{cha:observables}}

In the analysis of Hamiltonian (\ref{eq:EBHM}), we employ the Stochastic Series Expansion (SSE) code, a QMC algorithm from the ALPS  (Algorithms and Libraries for Physics
Simulations) project \cite{ALPS}. 
We mainly rely on three observables.  
First, we study the density, $\rho=\langle n_i \rangle$,  as a function of the chemical potential. Plateaus in the corresponding graph as a function of chemical potential indicate insulating phases, such as MI or CDW phases. 
The employed variant of QMC works in the grand-canonical ensemble, i.e., at fixed chemical potential. 
Discontinuous jumps in the density as a function of chemical potential signify regions of phase separation (PS) in the canonical phase diagrams. 
Namely, when the filling is fixed to a value which is not stable at any chemical potential, the system acquires the required filling only in the mean, by forming domain walls between two phases that are thermodynamically stable.

To distinguish not only different insulating phases (MI, CDW I, and CDW II), but also the superfluid (SF) and the supersolid phase (SS), we consider two other observables. 
These are the structure factor and the superfluid stiffness, which we analyze both as a function of density.  
The structure factor is defined as 
\begin{eqnarray}
  S(\mathbf{Q})=\left\langle\left\vert \sum\limits_{i=1}^{N}n_{i}e^{\imath \mathbf{Qr}_{i}} \right\vert^{2}\right\rangle/N^{2}\,. 
  \label{ST}
\end{eqnarray}

\noindent Here, $N$ denotes the number of lattice sites, and we focus on the wave vector ${\bf Q}=(\pi,\pi)$, which corresponds to a checkerboard modulation pattern. 
This observable has a peak when the particles are arranged in either of the CDW phases. 
This will help to distinguish the MI phase from the CDW phase, which cannot be done from the density graphs alone. For example, when a system is at unit filling, the structure factor is finite in the CDW II state, whereas it vanishes in a usual MI state. 

The other observable is the superfluid stiffness, which can be calculated from the winding numbers of the QMC code.  It is defined as
\begin{eqnarray}
  \rho_{s}=\frac{\langle W^{2} \rangle }{4 \beta},
  \label{SF}
\end{eqnarray}

\noindent where $W$ is the winding-number fluctuation of the world lines and $\beta$ is the inverse temperature (in this study $\beta=20$).  
This value shows what percent of the system is in a superfluid state. 
Taking superfluid stiffness and structure factor together, we can also identify the SS phase. The SS phase occurs when both 
superfluid stiffness and structure factor are non-zero. 
Note that, since PS regions do not correspond to stable grand-canonical phases as computed in the SSE QMC code, we cannot assign any values of observables for them. This is not necessary, however, since PS regions are already unambiguously identified by jumps in plots of density against chemical potential.

From these three observables (density, structure factor, and superfluid stiffness) we are now able to distinguish the most prominent phases that we are looking for.
These observables cannot, however, identify PSF phases, the signature of which is, as mentioned previously, a non-vanishing two-particle NN correlation function $\Phi_{i}$.  
In its current version, the QMC code provided in the ALPS library is not able to calculate these correlation functions. 
In order to extract this observable, the code would have to be written with a two-headed worm, which could then be analyzed in a similar way as the superfluid stiffness, but with the difference that the winding numbers would represent the flowing of pairs instead of single particles \cite{Bonnes11}. 
Fortunately, one can identify a dominant PSF order parameter using a different technique, namely by studying the density histograms of the QMC code. If these histograms show only even values of particles instead of a uniform distribution, this 
means that the bosons always pair up, indicating PSF behavior \cite{Bonnes11}. 

\section{Ground-state phase diagrams}

In this section, we present our QMC results for the ground-state phase diagram of Hamiltonian~(\ref{eq:EBHM}).
We focus on a two-dimensional square lattice with linear system sizes ranging from $L=8$ to $L=16$ (where $N=L\times L$). 
We present phase diagrams at two different values of the on-site repulsion ($U=20$ and $U=5$) for varying density and $T$ (and therefore for varying $V$, since $V=10|T|$).  
The two $U$ values are chosen in such a way that we can compare nearly hard-core like behavior, achieved at $U=20$, with soft-core behavior, for $U=5$.  
Further, at $U=20$ we can compare our data to known results of the usual EBHM, which was studied thoroughly in \cite{Sengupta05}. We compare phase diagrams obtained with and without density-dependent tunnelings. 
For simplicity and ease of comparison to \cite{Sengupta05}, we restrict our study to unit filling or less.
Furthermore, for a more detailed evaluation of these phase diagrams, we study a few cuts at representative parameter values.  

\begin{figure}[ht!]
\centering
\includegraphics[width=0.43\columnwidth]{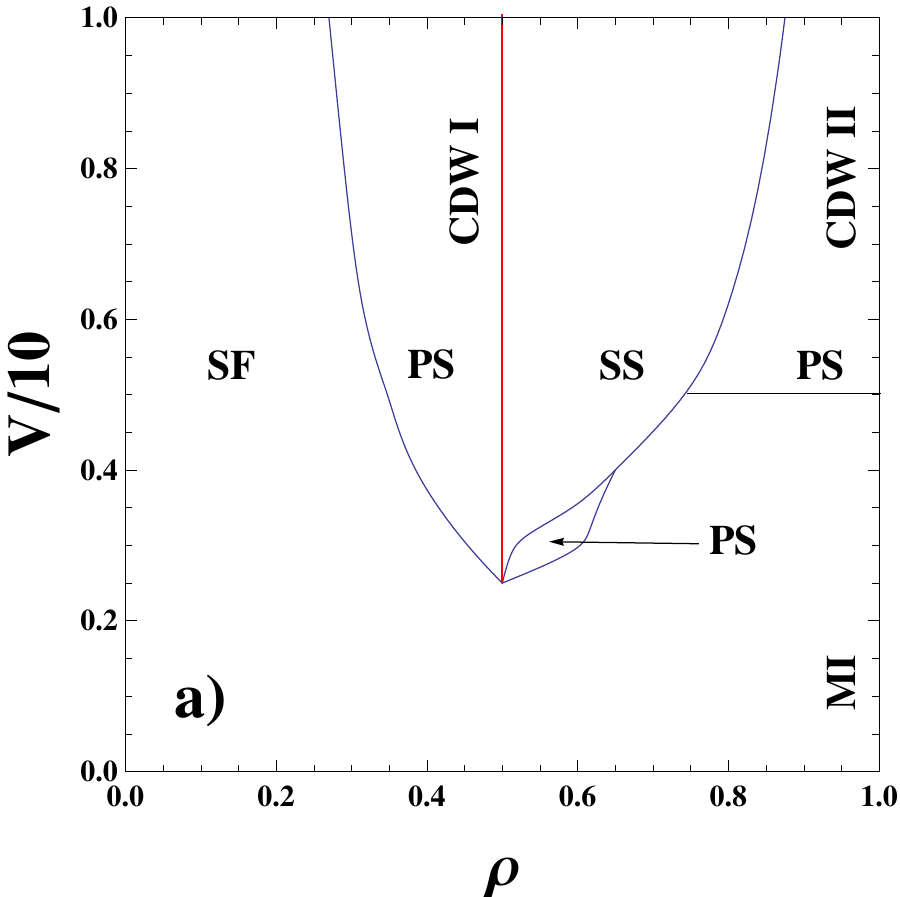}
\quad
\quad
\quad
\includegraphics[width=0.43\columnwidth]{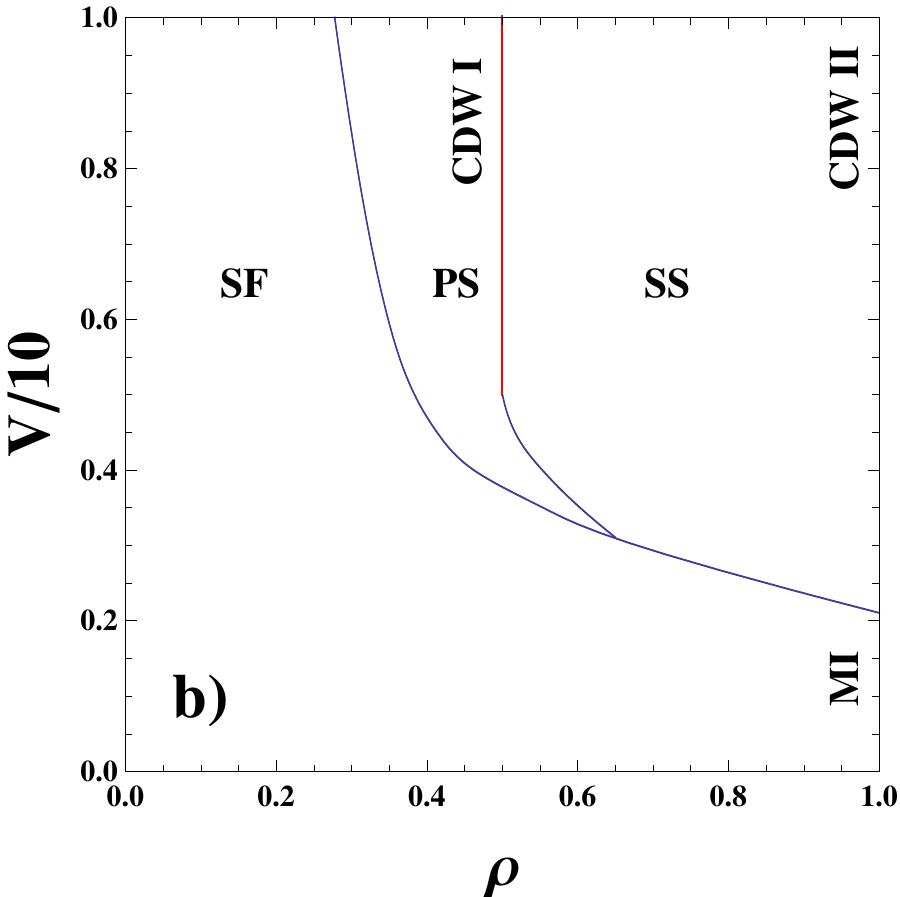}

\caption{The phase diagram in the $\rho-V$ parameter space without density-dependent tunneling term, $T=0$, for (a) $U = 20$ and (b) $U=5$. The energy unit is $t=1$. 
Panel a) reporoduces the results of \cite{Sengupta05}. 
The model contains various phases. The red solid line indicates the charge density wave (CDW I) at half filling; other phases present are the superfluid (SF), supersolid (SS), and at unit filling either Mott insulator (MI) or another charge density wave (CDW II); PS denotes phase-separated regions. 
When the on-site interaction becomes weaker, as shown in panel b), the SS phase becomes larger and PS regions disappear at filling larger than $1/2$.
}
\label{fig1-noT}
\end{figure}

\subsection{Phase diagrams at vanishing density-dependent tunneling}

We begin our analysis with phase diagrams of the regular EBHM, illustrated in Fig.~\ref{fig1-noT}. 
This provides an overview of the behavior of the considered systems under a more common Hamiltonian, which does not have a density-dependent term $T$. We consider the case of strong repulsion $U=20$, discussed previously in \cite{Sengupta05}, 
as well as softer interacting bosons with $U=5$, where up to 4 bosons are allowed per site.

\subsubsection{Phase diagram at strong on-site repulsion ($U=20$)}

For ease of comparison, and for later reference, Fig.~\ref{fig1-noT}a reproduces the phase diagram of $U=20$ that has been thoroughly investigated in \cite{Sengupta05}. It is well known that for $\rho < \frac{1}{2}$ there exist only two distinct regions, the SF phase and a PS region. 
For sufficiently low values of $V$, the system stays superfluid across the entire density range until unit filling, where it becomes a MI state. At half filling, a CDW I phase appears at a critical value of $V$, which in the present case lies around $V=2.5$. 
A system in a checkerboard phase (CDW I) can be doped by holes or particles. 
When it is doped with holes, these create domain walls and cause the system to phase separate, preventing the appearance of a SS phase.  In the case of hardcore bosons, this behavior would be mirrored for $\rho > \frac{1}{2}$, due to particle--hole symmetry. 
In the case of soft-core bosons, such particle--hole symmetry can break down.  
At sufficiently low $V$, a region of PS appears, and the system does present a hardcore-like behavior, but as the NN repulsion is increased this PS region disappears. 
Since now the particles can occupy either an empty or occupied site, it is no longer necessary for the domain walls to form and the system can move into a SS state. 
Moreover, at a certain value of $V$, upon increasing $\rho$ the SS phase is followed by a region of PS, instead of going into a SF phase and then becoming a MI. At unit filling, this PS region then changes to the CDW II phase, 
which is characterized by a checkerboard pattern consisting of an alternation of doubly-filled sites and empty ones. 

\begin{figure}[ht!]
\centering
\includegraphics[width=0.41\columnwidth]{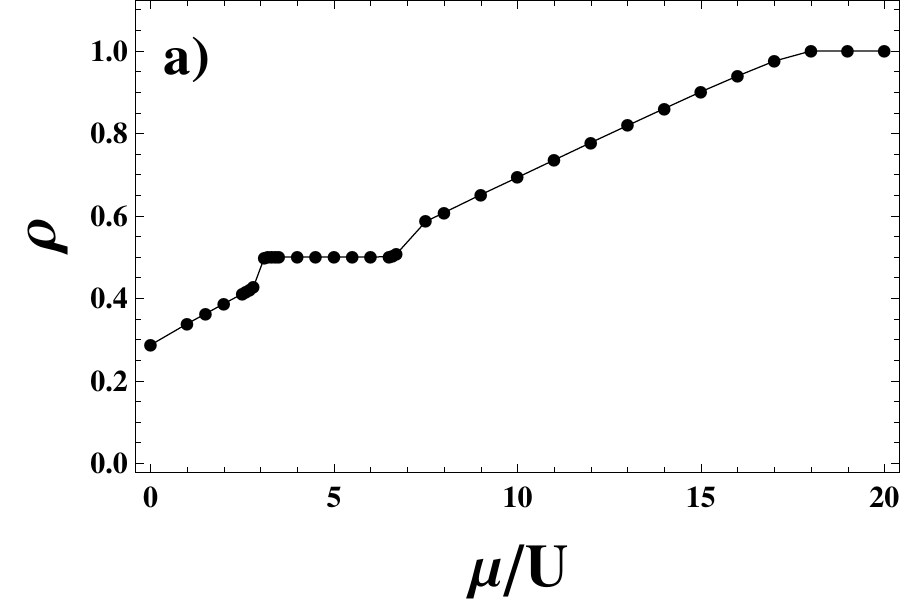}
\quad
\quad
\quad
\includegraphics[width=0.47\columnwidth]{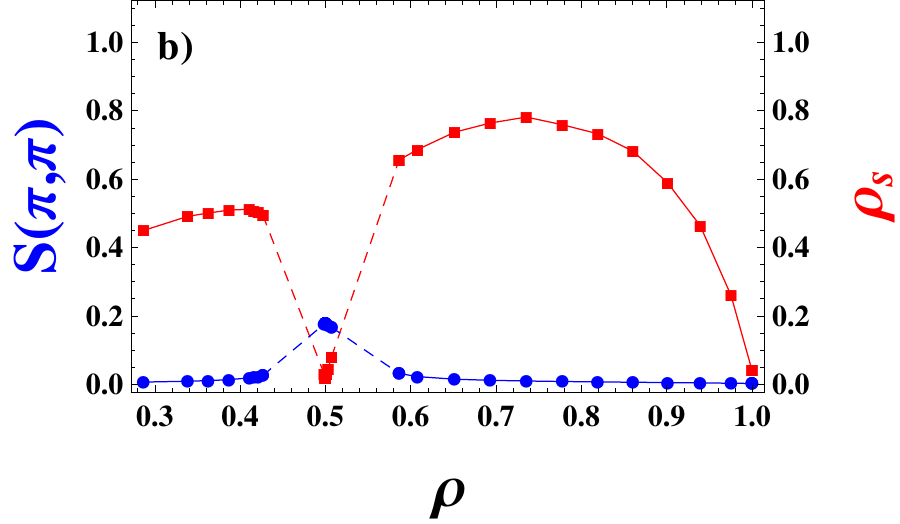}

\caption{Identification of different phases as exemplified for $U=20$ and $V=3.0$ based on the density in (a) as well as the
structure factor (blue circles) and the superfluid stiffness (red squares) shown versus density in (b).   
Plateaus in $\rho$ as a function of $\mu$ indicate incompressible crystal phases. Jumps denote phase separation (the densities that are jumped over do not correspond to thermodynamically stable phases). 
Moreover, a finite superfluid stiffness characterizes SF phases and a finite structure factor CDW order. When both are finite, the system is SS. 
}
\label{figU20}
\end{figure}

Figure~\ref{figU20} shows, for a fixed $V=3$, the observables described in Section~\ref{cha:observables} that we used to determine the various phases. 
The boson density as a function of the chemical potential displays clear plateaus, corresponding to gapped insulating phases (Fig.~\ref{figU20}a). 
As mentioned above, jumps in Fig.~\ref{figU20}a correspond to PS regions in Fig.~\ref{fig1-noT}. 
The structure factor and the superfluid stiffness are shown in Fig.~\ref{figU20}b. 
For low chemical potential (density) the system is in a SF state with non-zero superfluid stiffness and vanishing structure factor. At $\rho \approx 0.43$, the system phase separates, and there are no values for these observables. 
At half filling, when the system moves to CDW I phase, the structure factor becomes finite. There is a small region, roughly around $0.5 < \rho < 0.51$, where the system is in a SS phase -- here both the superfluid stiffness and the structure factor are non-zero. 
This phase is followed by a second region of PS that extends up to $\rho \approx 0.61$. At higher densities, a SF phase is observed up to unit filling, where a MI state follows, as revealed by vanishing superfluid density and structure factor.

\subsubsection{Phase diagram at moderate on-site repulsion ($U=5$)}

We now consider $U=5$, a case of weaker repulsion that has not been studied earlier. For the moment, we still retain $T=0$. 
The phase diagram Fig.~\ref{fig1-noT}b seems a bit simpler than for $U=20$.
Importantly, the particle-doped side now has to deal with much "softer" bosons allowing for multiple occupancy on any given site. 
The hole-doped side is much less affected since the on-site repulsion has a lesser influence on lower densities.
For weak NN repulsion $V$, the system stays SF across the entire range of densities from empty to unit filling and 
then goes into the MI state.  At $V\approx 2.3$ up to $V\approx 3.1$, the system goes directly from a SF phase into a SS phase, which ends at a CDW II phase at unit filling. At larger $V$, a PS region appears. The biggest difference between $U=20$ and $U=5$ cases appears for higher values of $V$, where the PS region at the particle-doped side disappears and the SS phase occupies the entire region between the CDW I at half filling and the CWD II at unit filling.

\begin{figure}[ht!]
\centering
\includegraphics[width=0.99\columnwidth]{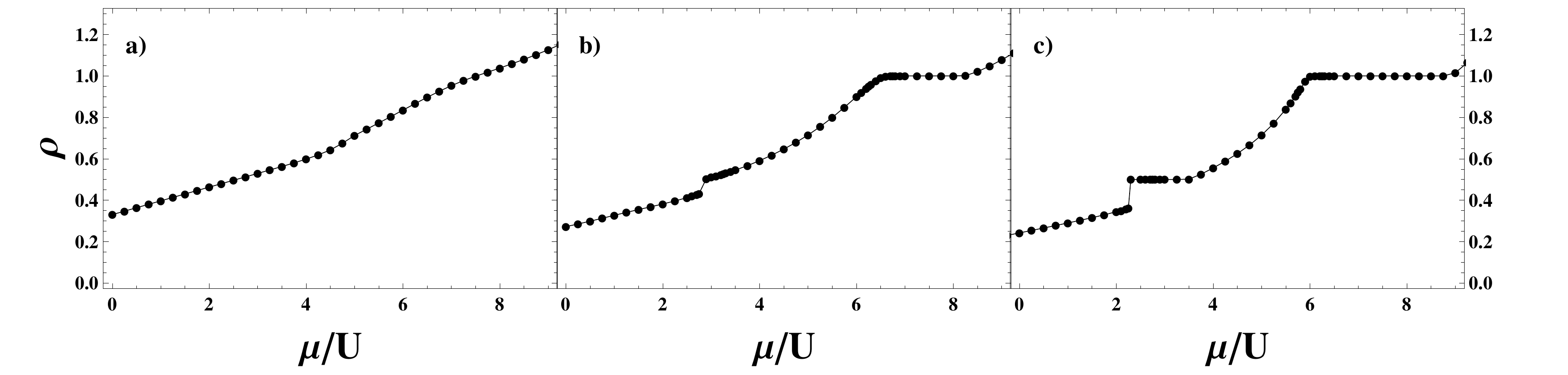}
\includegraphics[width=0.99\columnwidth]{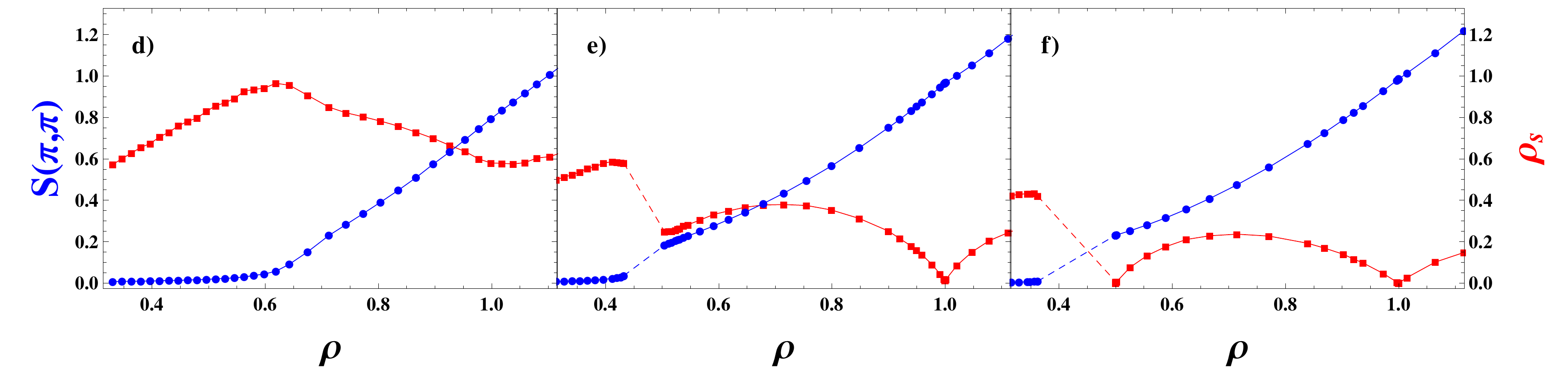}

\caption{Top row: density versus chemical potential for $U = 5$ and different values of $V$: $V=3.0$, $V=4.5$ and $V=6.0$ (left to right). The bottom row shows the corresponding structure factor (blue circles) and the superfluid stiffness (red squares).
}
\label{figU5}
\end{figure}

The different transitions are revealed by a slices through the phase diagram at fixed $V$ (exemplified for a few values in Fig.~\ref{figU5}).  At $V=3.0$, the density is strictly increasing across the entire range of $\mu$ (Fig.~\ref{figU5}a). 
Notice, however, a change of the slope around $\mu/U=4$, corresponding to $\rho=0.6$. 
As seen in Fig.~\ref{figU5}d, the structure factor starts to rise in a similar parameter range, namely around $\rho=0.65$. 
At the same time, the superfluid stiffness only has a peak at $\rho=0.65$, but remains finite for all values of $\mu$ considered. 
Therefore, the increase of the structure factor is a clear sign of a second-order transition from a SF to a SS. 
Also, since the structure factor does not drop back to zero at $\rho=1$, the unit-filling phase will be a CDW II and not a MI.

The next slice is taken at $V=4.5$, where the state changes from SF to PS to SS without 
ever settling into the CDW I phase at half filling. 
In the density graph, Fig.~\ref{figU5}b, we can see a small jump that bypasses 
$\rho=\frac{1}{2}$.  This explains why the CDW I phase does not appear at this value of $V$. The SF at small $\rho$ is identified by a non-zero superfluid fraction and vanishing structure factor (Fig.~\ref{figU5}e). 
This phase is followed by the PS region from $\rho \approx 0.435$ to $\rho \approx 0.51$. 
At higher densities, a SS state appears as characterized by non-zero structure factor and superfluid stiffness.  
Finally, the system settles into the CDW II state at unit filling.

The last slice at $V=6.0$ is similar to the previous one at $V=4.5$ with one major difference, the appearance of the CDW I phase at half filling. As before, we can see a jump (this time slightly larger) in the density, Fig.~\ref{figU5}c, but now it is followed by a plateau that signifies the CDW I phase. In Fig.~\ref{figU5}f, we see again the three distinct phases, SF up to $\rho \approx 0.35$, then a region of PS up to $\rho=0.5$, and from half filling to unit filling there is the SS phase, once again ending in the CDW II state.
 
\subsection{Phase diagrams at finite density dependent tunneling}

As we have seen in the previous section, the phase diagram of the EBHM at vanishing $T$ displays a large variety of phases: MI, CDW, SF, and SS. Additionally, there are various regions of phase separation, some of which (the ones at filling larger than 1/2) disappear with decreasing on-site repulsion of the bosons. In this section, we study how this phase diagram of the usual EBHM is changed by the density-dependent hopping.  

\begin{figure}[ht!]
\centering
\includegraphics[width=0.43\columnwidth]{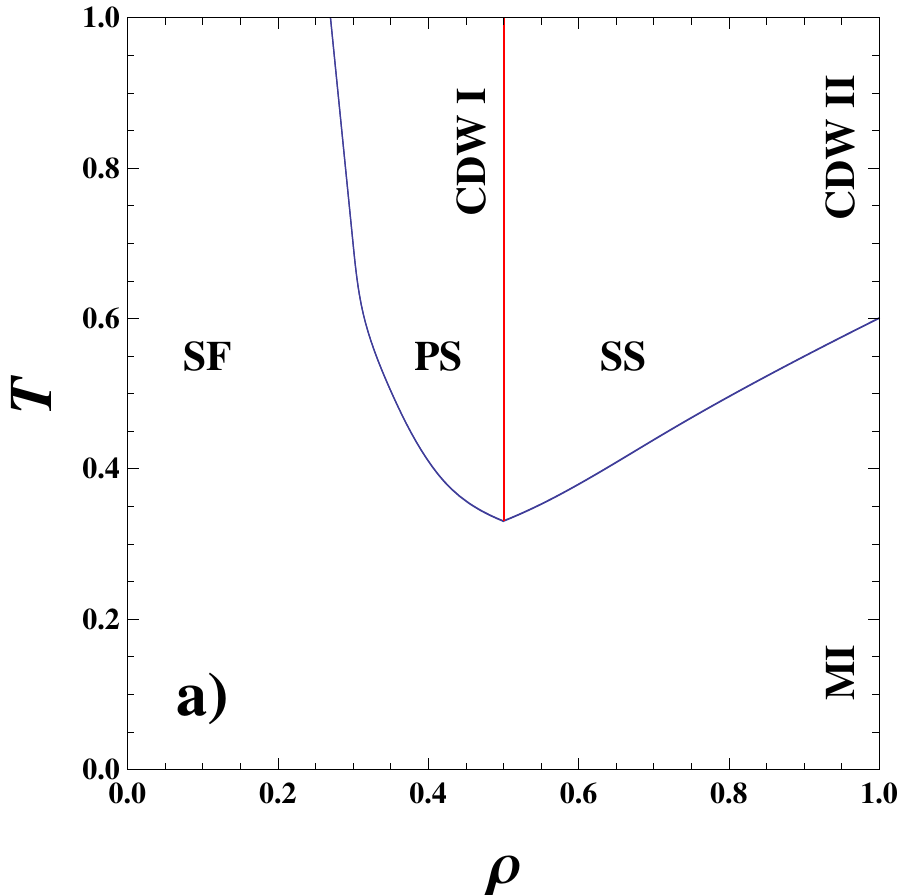}
\quad
\quad
\quad
\includegraphics[width=0.43\columnwidth]{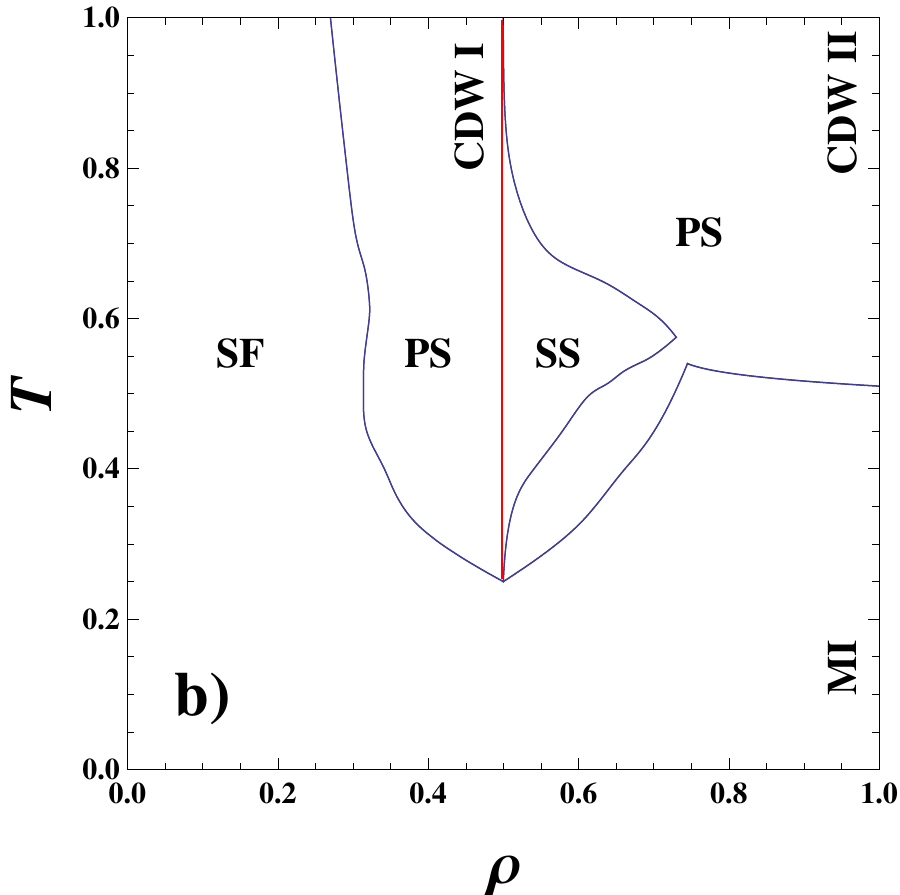}

\caption{The phase diagrams for $U = 20$ at finite $T$ (with $V=10|T|$, and $t=1$ the unit of energy).
(a) If $t$ and $T$ are of the same sign, the relative importance of interactions decreases, leading to the disappearance of PS phases at greater than half filling. Compared to the $T=0$ cases presented in Fig.~\ref{fig1-noT}, this phase diagram resembles more the case $U=5$ than $U=20$. 
(b) If $T$ and $t$ compete due to opposite sign, the relative importance of interactions is enhanced, increasing the PS regions.  In fact the two separate regions of PS in Fig.~\ref{fig1-noT}b increase to the point of overlapping.}
\label{figU20T}
\end{figure}

\subsubsection{Phase diagram at strong on-site repulsion}

The first case we study is $U=20$ when the two tunneling amplitudes $t$ and $T$ have the same sign.  
Comparison of Fig.~\ref{figU20T} with Fig.~\ref{figU20} shows that in the presence of density-dependent tunneling the PS region at low $V$ values has disappeared and there is 
no PS region between the SS and CDW II phases. One can explain this behavior by the increase in the total hopping due to the additional tunneling term $T$. 
Thus, the on-site repulsion $U$ behaves as if effectively rescaled to smaller values. Similar arguments explain the shift of the point where the $\rho = \frac{1}{2}$ plateau first appears and, therefore, the CDW I phase moves from $V \approx 2.5$ (with $T=0$, Fig.~\ref{figU20}) to $V  \approx 3.5$ (Fig.~\ref{figU20T}).
As a consequence, the phase diagram at $U=20$ with $t$ and $T$ of the same sign looks very similar to the one at $U=5$ with vanishing $T$.

The behavior in the $U=20$ phase diagram becomes more interesting when the two tunneling terms compete due to opposite signs, $T<0$. The phase diagram is presented in  Fig.~\ref{figU20T}b. The CDW I phase now starts at a lower value of $|T|$ than in the previously discussed case.  Similarly, the region of PS at the lower values of $|T|$ (and thus $V$) now becomes much larger.  This shows that the system has a hardcore behavior for a larger range of parameters. Additionally, the SS region diminishes and finally disappears as $V$ gets larger. 
These findings can be explained through the competition between $t$ and $T$, which decreases the effective, overall tunneling strength. This decrease can alternatively be seen as an effective relative increase of the interaction parameters $U$ and $V$. As a result, the hard-core behavior of the system becomes more pronounced, and the PS regions become more important.

\begin{figure}[ht!]
\centering
\includegraphics[width=0.99\columnwidth]{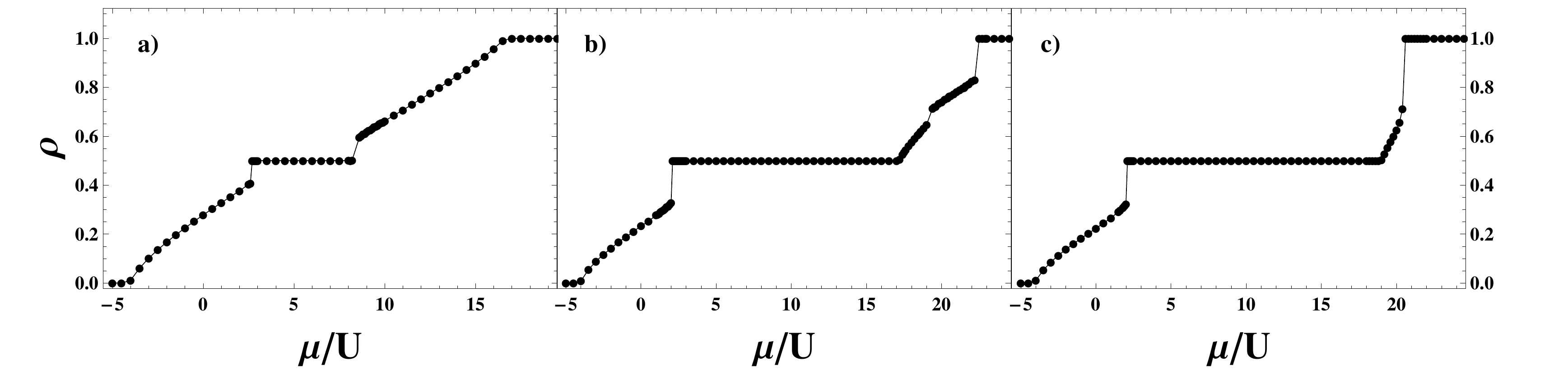}
\includegraphics[width=0.99\columnwidth]{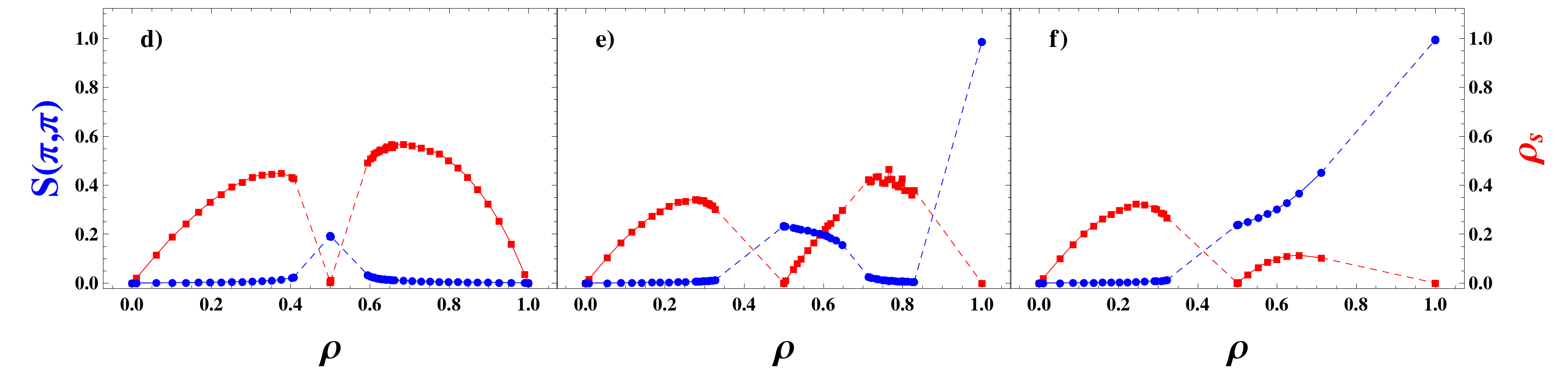}

\caption{Density graphs of $U = 20$ for $T=-0.3$, $T=-0.5$ and $T=-0.6$ (left to right).
Structure factor (blue circles) and superfluid stiffness (red squares) graphs for $U = 20$ at $T=0.3$, $T=0.5$ and $T=0.6$ (left to right)
}
\label{cutU20T}
\end{figure}

The observed phases may again be analyzed in detail via the cuts at fixed $V$ ($T$), presented in Fig.~\ref{cutU20T}. 
The first slice we present is for $T=-0.3$ ($V=3.0$). As seen in Fig.~\ref{cutU20T}a, the plateau at half filling --- a CDW I, as indicated by the finite structure factor, Fig.~\ref{cutU20T}d --- is surrounded by discontinuities in the density, thus implying regions of PS. These are surrounded by SF phases, with a MI appearing at unit filling.  

The next slice cuts through the phase  diagram at $T=-0.52$ ($V=5.2$) and this time shows also a region of the SS phase for densities just above half filling (Fig.~\ref{cutU20T}e). This SS may also be observed in the density plot, Fig.~\ref{cutU20T}b: Above half filling, there is a small interval of steady increase before a discontinuity occurs around $\rho=0.65$.  
After this PS region, there is a small region where the system becomes superfluid before once again phase separating.  At unit filling, the system finally transitions into a CDW II phase.
Below half filling, another jump in the density indicates yet another PS. 

The final cut is taken at $T=-0.6$ ($V=6.0$).  Again, at low densities the system starts in a SF phase and then jumps through a region
of PS to reach the CDW I phase at half filling. For higher densities, the system first enters a SS phase, and around $\rho=0.72$ a transition to PS occurs. This time, the system ends in the CDW II phase when unit filling is reached.

\begin{figure}[ht!]
\centering

\includegraphics[width=0.43\columnwidth]{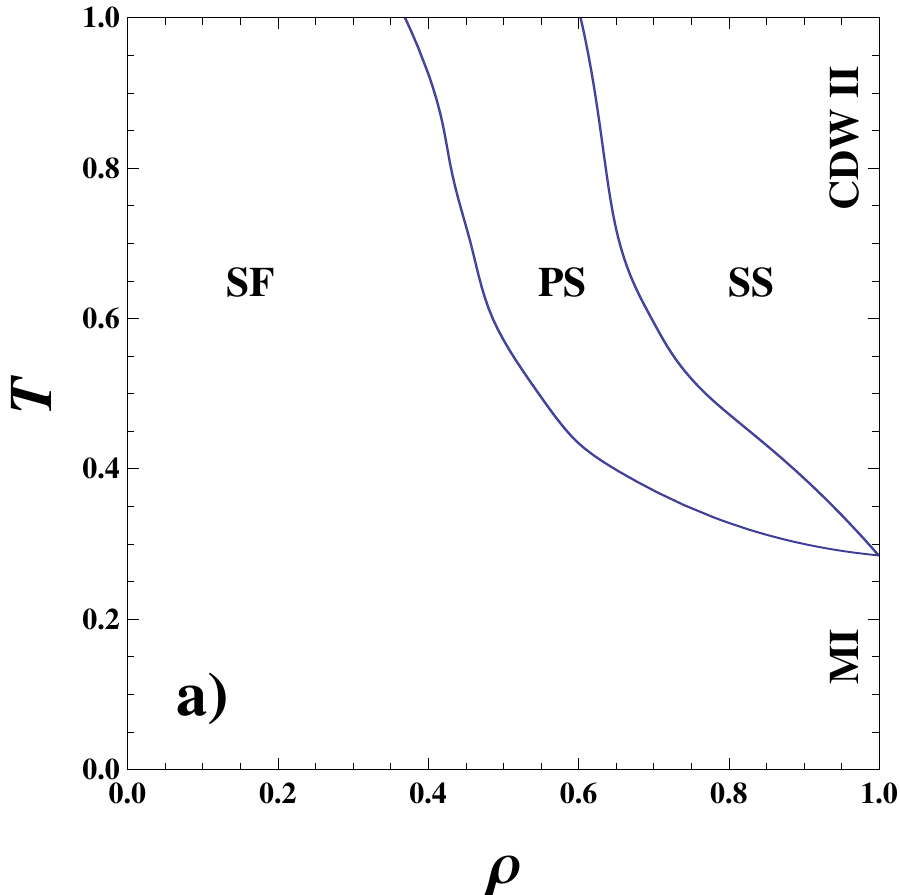}
\quad
\quad
\quad
\includegraphics[width=0.43\columnwidth]{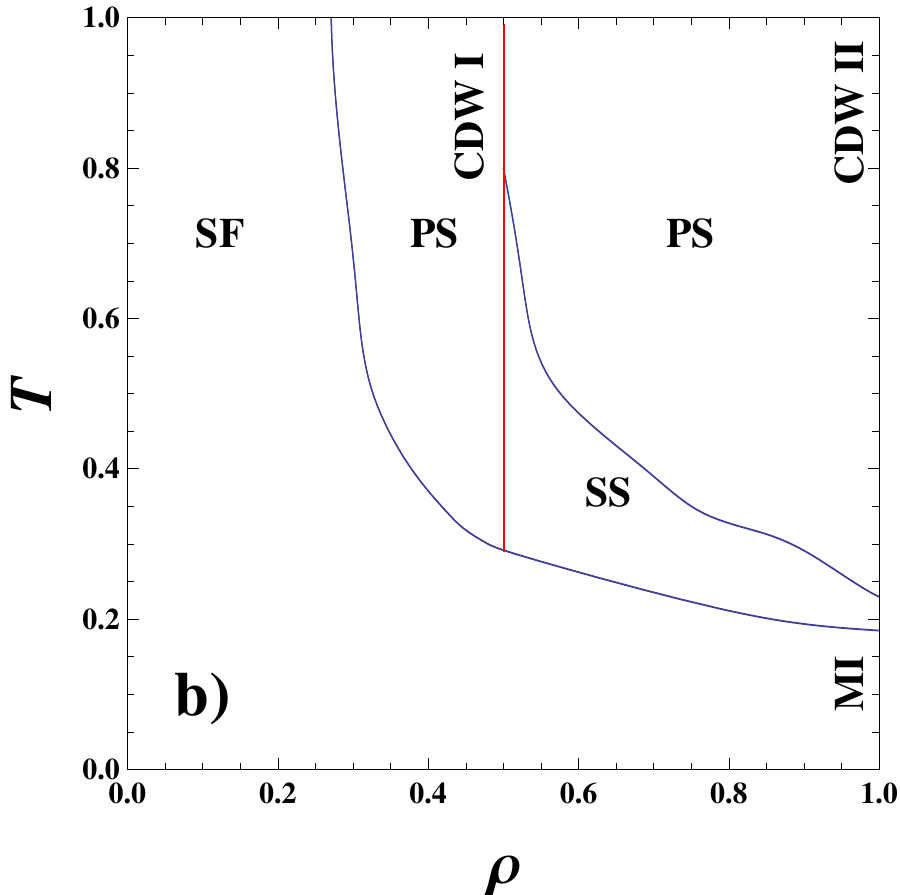}

\caption{Phase diagrams for $U = 5$ and finite $T$. 
(a) If $T$ and $t$ have the same sign, the relative strength of tunneling is strongly increased with respect to the interactions. As a consequence, the CDW I phase has disappeared completely from this phase diagram. 
(b) When $T$ and $t$ are of opposite sign, the role of interactions is enhanced, leading to increased PS regions and again the CDW I phase is present.}
\label{figU5T}
\end{figure}

\subsubsection{Moderate on-site repulsion ($U=5$)}

In the previous section, we saw that the additional density-dependent tunneling term $T$ can increase or decrease the effective importance of the interactions $U$ and $V$, depending whether it competes with or supports the single-particle tunneling $t$. 
In this section, we study this effect  for  weaker on-site interaction $U=5$.  
The corresponding phase diagrams are presented in Fig.~\ref{figU5T}. 

The positive-$T$ diagram reveals that the CDW I phase, present for $T=0$, disappears completely (Fig.~\ref{figU5T}a).  This means that at no point does there exist a plateau in the density graphs at 
$\rho=\frac{1}{2}$. Instead, a discontinuity bypasses half filling altogether. The rest of the behavior is
rather similar to the system without the density-dependent term. There are still only  three phases below unit filling, i.e., the SF phase at low densities and $T$ (and therefore at low $V$), the PS region near half filling for larger $T$, and finally
the SS phase for still higher $T$ and larger densities. 
As can be expected, when the SF phase persists through the entire range of densities, the system ends in a MI state at unit filling. Instead, when the system at fixed $V$ passes through the SS state, the final phase at unit filling is, as before, the CDW II phase.

Consider now the phase diagram of a system with $U=5$ when the tunneling terms have opposite signs (Fig.~\ref{figU5T}b). Here, contrary to the case of positive $T$, the CDW I exists at half filling.
This indicates that the relative importance of the effective total tunneling is suppressed for $T<0$. Moreover, now a second region of PS appears above half filling. As a result, for $T\lesssim -0.55$ there is no stable phase with a density between the CDW I and the CDW II. 

\begin{figure}[ht!]
\centering
\includegraphics[width=0.99\columnwidth]{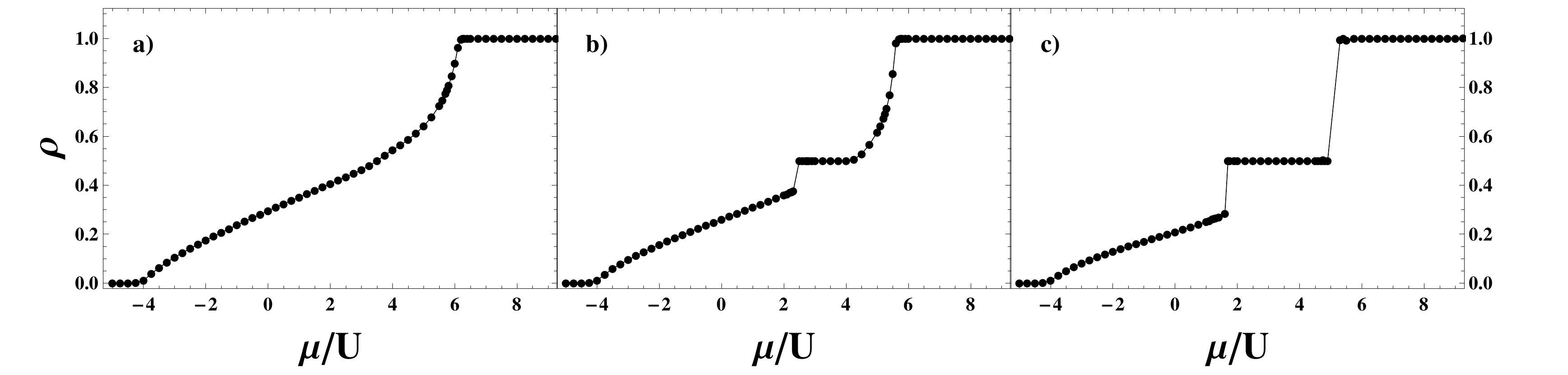}
\includegraphics[width=0.99\columnwidth]{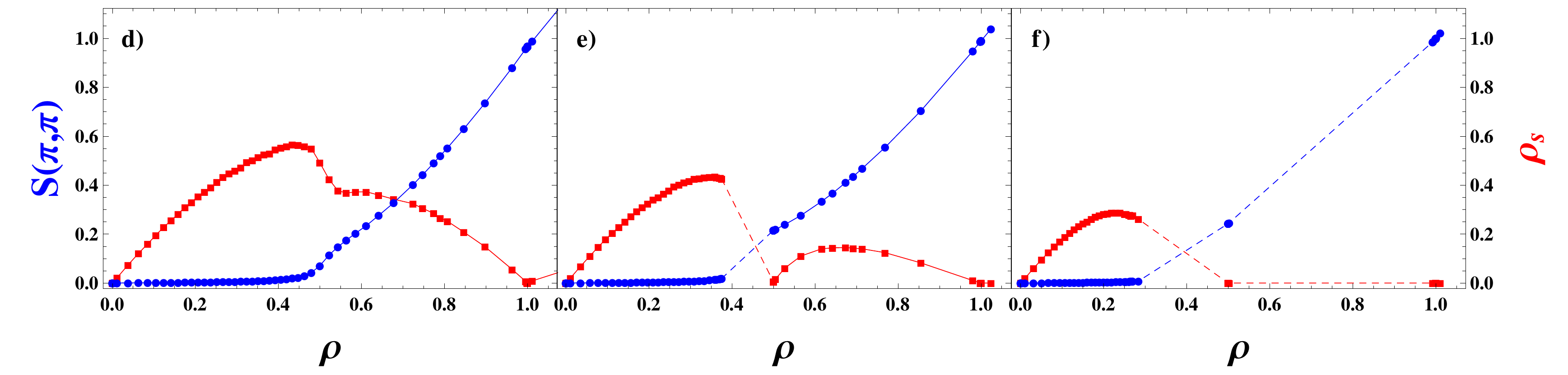}

\caption{Top row: Density graphs for $U = 5$ and $T=-0.3$, $T=-0.4$ and $T=-0.8$ (left to right). The bottom row shows the structure factor  (blue circles) and the superfluid stiffness (red squares) for the same parameters}
\label{cutT5}
\end{figure}

These observations about the phase diagram are supported by an analysis of cuts at a few chosen values of $T$ (and thus $V$), see Fig.~\ref{cutT5}.
At $T=-0.3$ ($V=3.0$), one observes a smooth density increase all the way until unit filling, where a plateau appears (Fig.~\ref{cutT5}a). The structure factor starts increasing near half filling, indicating the transition from the SF  to the SS phase (Fig.~\ref{cutT5}d); at unit filling, the system lands in the CDW II phase.

A cut at the slightly higher absolute value $T=-0.4$ ($V=4.0$) reveals a plateau at half filling (CDW I) and a second one  at unit filling (CDW II). Comparing the density plot (Fig.~\ref{cutT5}b) with those of superfluid stiffness and structure factor (Fig.~\ref{cutT5}e), we see that upon increasing the chemical potential the SF phase appears at low densities, followed by the PS which transitions into the CDW I at half filling. For higher densities, there is a region of SS, where both structure factor and superfluid stiffness are non-zero. 
Finally, there is the jump caused by the PS region directly to the CDW II phase at unit filling.

Let us finally consider stronger density-dependent tunneling $T$ and inter-site repulsion $V$, namely $T=-0.8$ ($V=8.0$).  
Below half filling, the density gradually increases up to the value of $\rho\approx0.27$ and then jumps to the CDW I phase (Fig.~\ref{cutT5}c). After this phase, the density behaves step-like, jumping directly into the CDW II phase at $\rho=1$. 
This behavior is seen clearly in the data presented in Fig.~\ref{cutT5}f, where the SF phase for low densities is followed by two distinct regions of PS.  
These regions are only interrupted by the CDW I phase at half filling and the CDW II phase at full filling.

As these results show, for the lower on-site interaction $U=5$, the density-dependent term $T$ does not change much the overall behavior of the phase diagram if it has the same sign as the single-particle tunneling $t$. Instead, if the two tunneling terms have opposite sign, a large part of the SS phase disappears into a phase-separated region, due to the increased relative importance of the interaction terms.

\section{Conclusion}

In summary, we have studied the extended Bose--Hubbard model on a square lattice with additional terms coming from density-dependent tunnelings. Taking these terms into account is relevant for experiments on ultracold dipolar molecules in optical lattices. 
The competition between the density-dependent tunneling, a standard single-particle hopping, finite on-site repulsion, and nearest-neighbor repulsion gives rise to a rich phase diagram of the system. 

Specifically, as has been found previously \cite{Sengupta05}, at large on-site repulsion and without density-dependent tunneling, there are Mott-insulator, charge-density wave, superfluid, and supersolid phases, as well as phase-separated parameter regions. 
Depending on the parameter strengths, this phase diagram undergoes considerable deformations. 
If either we reduce on-site repulsion or introduce density-dependent tunnelings that have the same sign as the single-particle hopping, some of the phase-separated regions disappear. 
Remarkably, if we introduce both of these effects simultaneously, additionally the charge density wave at half filling disappears.
In this case of same-sign tunnelings, both hopping processes act constructively producing an effective larger tunneling, or respectively, weaker interactions. 

We have also studied the phase diagram when the density-dependent tunneling and single-particle hopping compete due to the their signs being opposite. Due to this competition, the relative importance of interaction terms is enhanced. In this case, the most striking effect is the disappearance of the supersolid into a phase-separated region. This occurs on the particle-doped side of the half-filling charge density wave, and at strong $V$. Contrary to similar models in one dimension \cite{Sowinski}, we find no indications for pair-superfluid behavior for all considered parameter values. 

Besides a theoretical interest in understanding how density-dependent tunneling terms change phase diagrams of extended Bose--Hubbard models, our findings will help to determine where one may expect exotic phases in experiments with ultracold dipolar molecules in optical lattices. 

 \section*{Acknowledgements}

This work was supported by the International PhD Projects
Programme of the Foundation for Polish Science within the European
Regional Development Fund of the European Union, agreement no.
MPD/2009/6. We acknowledge financial support from Spanish
Government Grant TOQATA (FIS2008-01236) and Consolider Ingenio EU IP SIQS, ERC Advanced
Grant QUAGATUA, CatalunyaCaixa, Alexander von Humboldt Foundation and Hamburg
Theory Award. M.M.\ and J.Z.\ thank Lluis Torner, Susana Horv\'ath and all ICFO personnel for
hospitality. O.D.\ and J.Z.\ acknowledge support from Polish
National Center for Science project No. DEC-2012/04/A/ST2/00088.
P.H.\ acknowledges support from the Austrian Science Fund (SFB F40 FOQUS) and the Marie Curie Initial Training Network COHERENCE.

\end{document}